\documentclass{IEEE_lsens}

\usepackage{textcomp}

\usepackage[noadjust]{cite}

\usepackage[T1]{fontenc}
\usepackage{amsmath}
\usepackage[cmintegrals]{newtxmath}

\usepackage{bm}

\usepackage{array}

\usepackage{url}

\usepackage{graphicx}
\usepackage{dcolumn}
\usepackage{physics}
\usepackage{siunitx}

\DeclareSIUnit{\Molar}{M} 

\newcommand\MYhyperrefoptions{hypertexnames=true, bookmarks=true,
bookmarksnumbered=true, pdfpagemode={UseOutlines}, plainpages=false,
pdfpagelabels=true, colorlinks=true, linkcolor={black},
citecolor={black}, urlcolor={black}}

\usepackage[\MYhyperrefoptions,breaklinks=true,pdftex]{hyperref}

\providecommand{\hypersetup}[1]{\relax}

\hypersetup{pdftitle={Infrasound_Detection},
pdfsubject={Infrasound detection using polymer networks in liquid films.},
pdfauthor={Maarten Mittmann, Carsten Habenicht, Matthias Bornitz, Indraneel Sen, Hans Kleemann, and Karl Leo},
pdfkeywords={Sensor phenomena, electrochemical transducer, fiber network, free-standing liquid film, infrasound detection}}

\begin{document}

\markboth{Vol.~1, No.~3, July~2017}{0000000}

\IEEELSENSarticlesubject{Sensor phenomena}

\title{Infrasound Detection Using Polymer Networks in Liquid Films}

\author{\IEEEauthorblockN{Maarten~Mittmann\IEEEauthorrefmark{1}, Carsten~Habenicht\IEEEauthorrefmark{1},
Matthias~Bornitz\IEEEauthorrefmark{2}, Indraneel~Sen\IEEEauthorrefmark{3}, Hans~Kleemann\IEEEauthorrefmark{1}, and Karl~Leo\IEEEauthorrefmark{1}}
\IEEEauthorblockA{\IEEEauthorrefmark{1}Dresden Integrated Center for Applied Physics and Photonic Materials (IAPP) and Institute for Applied Physics, Technische Universität Dresden 01062 Dresden, Germany\\
\IEEEauthorrefmark{2}Medical Faculty of the Technische Universität Dresden, Department of Otorhinolaryngology, Head and Neck Surgery, Ear Research Center Dresden (ERCD), Dresden 01307, Germany\\
\IEEEauthorrefmark{3}Wasabi Innovations Ltd, Sofia 1113, Bulgaria\\}%
\thanks{Corresponding author: K. Leo (e-mail: karl.leo@iapp.de).\protect\\ 
M. Mittmann and C. Habenicht contributed equally.\protect\\
This project has received funding from the European Union’s research \& innovation program FETOPEN H2020-EU.1.2.1 under GA ID: 899205.}
\thanks{Associate Editor: }%
\thanks{Digital Object Identifier 10.1109/LSENS.2017.0000000}}

\IEEELSENSmanuscriptreceived{Manuscript received xxx; revised xxx; accepted xxx. Date of publication xxx; date of current version xxx.}

\IEEEtitleabstractindextext{%
\begin{abstract}
Recording and analyzing infrasound signals is essential to study natural phenomena such as earthquakes, weather, or avalanches, but also has practical importance in aviation industry, optimization of wind turbines, and many more. However, the detection of faint infrasound signals is still a significant challenge as the transducers (e.g., mechanical, optical, piezoelectric) are either difficult to integrate or do not provide sufficient sensitivity. Here, we propose an alternative principle to detecting infrasound which is based on a free-standing liquid film covered with a polymeric organic mixed ionic-electronic conductor (OMIEC). This polymer is capable of conducting ions as well as electrons/holes and serves as a direct electronic infrasound transducer due to the sensitivity of OMIECs to the local ion concentration in the liquid.
We specifically address the detection of acoustic excitations within the infrasound regime and the dependency of the networks reaction to these stimuli on its conductive properties, mainly its impedance spectrum. 
The resulting sensor has a sensitivity of $\SI{613}{\micro\volt\per\pascal}$ and a power consumption of about $\SI{340}{\nano\watt}$ at $\SI{100}{\milli\hertz}$, which puts these systems well within the range of commercial infrasound detectors, while offering significant advantages in terms of device complexity and integration.
\end{abstract}

}


\maketitle

\section{\label{sec:level1}Introduction}
The term infrasound typically refers to pressure variations at frequencies from around $\SI{20}{\hertz}$ down to a few mHz which is below the human audible spectrum \cite{HupeEarthSystemScienceData2022}. Infrasound is generated by a variety of sources, including weather events such as hurricanes \cite{ListowskiRemoteSensing2022}, volcanic eruptions \cite{DabrowaEarthandPlanetaryScienceLetters2011}, earthquakes \cite{Watada2006GRL}, ocean waves \cite{DeCarloGeophysicalJournalInternational2020}, and avalanches \cite{MayerColdRegionsScienceandTechnology2020}.
Moreover, infrasound can also be generated by human activity such as spacecrafts \cite{PilgerGeophysicalResearchLetters2021}, vehicles \cite{HaddadSAETransactions1989}, wind turbines \cite{KampAcousticsAustralia2018}, and nuclear tests \cite{Marty2019,HupeEarthSystemScienceData2022}.
Such infrasound waves can travel over long distances in the atmosphere without strong attenuation \cite{ListowskiRemoteSensing2022,HupeEarthSystemScienceData2022}. Hence, they can be used to study the types of events mentioned above with dedicated infrasound sensors or networks of sensors \cite{HupeEarthSystemScienceData2022, Marty2019}. 

Various types of sensors are in use to detect infrasound excitations, commonly grouped into sensors that measure differential and absolute pressure differences.
One way to determine absolute pressure differences is based on the principle of an aneroid capsule \cite{ Ponceau2010}. Such devices consist of a sealed elastic container with lower internal pressure than ambient pressure whose compression is counterbalanced mechanically and changes due to infrasound are converted to a corresponding voltage signal via a transducer.
Differential infrasound sensors (i.e. infrasound microphone) \cite{Ponceau2010} utilize a reference volume separated from ambient pressure by a bellow or diaphragm. A small leak in said volume allows an adjustment to ambient pressure over time. However, it prevents faster pressure changes on the infrasound scale or higher from entering the cavity permitting detection of the spatial displacement of the sensor by a transducer, such as magnet and coil velocity transducers \cite{AsmarTheJournaloftheAcousticalSocietyofAmerica2018}, linear variable differential transformers \cite{Fraden2016}, quartz crystal resonators \cite{Watada2006GRL}, or optical transducers \cite{KortschinskiJournalofPhysicsEScientificInstruments1971}.

Here, we explore an electronically active, free-standing liquid film containing a poly(3,4-ethylenedioxythiophene) (PEDOT) network as an infrasound sensor operating at frequencies below $\SI{1}{\hertz}$.
PEDOT is a conducting polymer that can be grown to form a fiber network (FN) in a solution to which an alternating potential is applied. The work shows that a current flowing through the fibers can be modulated by low-frequency acoustic or seismic excitations. The resulting pressure fluctuations are linearly converted by the system into an electrical signal. Thus, an acoustic sensor based on a FN in a liquid film would not require a mechanical transducer, but only a signal amplifier, which significantly reduces the complexity of integration.

\section{\label{sec:Methods}Methods}
\begin{figure}[!t]
  \centering
  \includegraphics[width=0.48\textwidth]{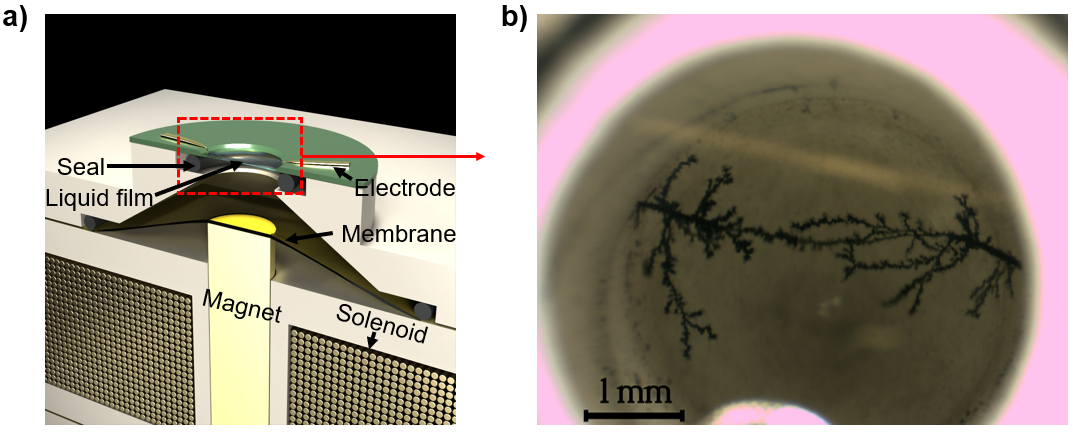}
  \caption{a) Schematic of the setup with built-in infrasound generator. b) A PEDOT:PF$_6$ FN grown by the authors in an older setup without the built-in infrasound generator. This picture only serves the purpose of showing what such FNs look like, the ones grown for the actual experiments have a lower fiber thickness (cf. Fig. \ref{fig:Vibrometer}b).}
  \label{fig:OscillatorSchematic}
\end{figure}

Experiments are performed on free-standing liquid films made from a solution of sulfolane, $\SI{1}{\milli\Molar}$ tetrabutylammonium hexafluorophosphate (TBAPF$_6$) and $\SI{50}{\milli\Molar}$ 3,4-ethylenedioxythiophene (EDOT). They have a diameter of $\SI{5}{\milli\meter}$ and a thickness of roughly $\SI{1}{\milli\meter}$ and are held by a $\SI{25}{\milli\meter}$\,x\,$\SI{25}{\milli\meter}$\,x\,$\SI{1.5}{\milli\meter}$ fiber-reinforced epoxy plate with a $\SI{5}{\milli\meter}$-diameter hole in the center (see Figure \ref{fig:OscillatorSchematic}a). The film is created at one end of a glass pipe with an inner diameter of $\SI{5}{\milli\meter}$ by dipping it in the solution and subsequently transferring it to the epoxy frame by placing this end of the pipe briefly on the edge of the hole in the frame. The central hole is tapered so the height of the inner wall touching the meniscus is around $\SI{500}{\micro\meter}$. This yields a high contact area between the frame and liquid film, which is then held in place by adhesion forces. Even at high oscillation amplitudes, these forces are enough to keep the film stable, due to its high viscosity and boiling point (stability of liquid film in frame: at least several days). Two $\SI{25}{\micro\meter}$ thick gold wires are positioned so that one end of each wire reaches into the central hole on opposite sides. During the experiment the tips of the gold wires stay immersed in the film.

PEDOT:PF$_6$ FNs \cite{Ciccone2022} are then grown from these wires by applying sinusoidal potentials with peak-to-peak amplitudes of $\SI{25}{\volt}\,-\,\SI{50}{\volt}$ at frequencies of $\SI{10}{\hertz}\,-\,\SI{20}{\hertz}$ to the wires using an Agilent 33622A waveform generator with a Falco Systems WMA-300 high voltage amplifier. The impedance of the FNs in the liquid films is measured via the gold wires using an Autolab PGSTAT128N potentiostat/galvanostat. Moreover, potentials of various shapes (DC, sinusoidal, step) and amplitudes are applied to the FNs with an Agilent 33622A waveform generator while the resulting currents flowing through the FNs are recorded with a FEMTO DLPCA-200 transimpedance amplifier and a Rohde \& Schwarz HMO3004 oscilloscope. Those measurements are carried out with and without acoustic or seismic excitations of the films containing the PEDOT:PF$_6$ FNs. The amplifier has an integrated $\SI{10}{\hertz}$ low-pass filter, which is used in all measurements for noise reduction. 

\begin{figure}[!t]
\centering
\includegraphics[width=0.48\textwidth]{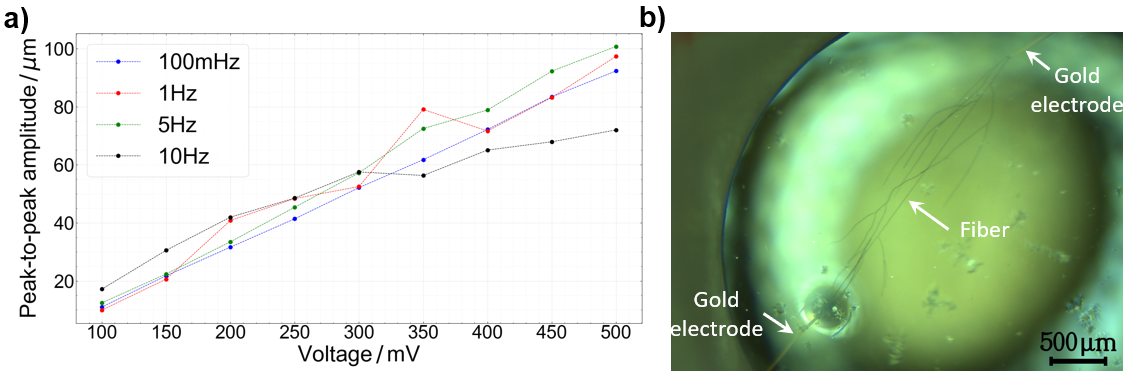}
\caption{a) Dependency of the peak-to-peak amplitudes of a free-standing liquid film on the voltage applied to the oscillator. The measurement is performed at four different excitation frequencies. b) Microscope image of a FN grown in a free-standing sulfolane film with \SI{50}{\milli\Molar} EDOT and \SI{1}{\milli\Molar} TBAPF$_6$ at a frequency of \SI{15}{\hertz}.}
\label{fig:Vibrometer}
\end{figure}

A seismic excitation is simulated by dropping an object with a mass of $\SI{70}{\gram}$ from a height of $\SI{15}{\centi\meter}$ on the setup table while a DC potential is applied and the current through the PEDOT:PF$_6$ FN is measured as described above. The applied acoustic excitations range from $\SI{0.1}{\hertz}$ to $\SI{50}{\hertz}$ and are generated by a home-made infrasound generator. This device consists of a solenoid with a bar magnet inside, which is connected to a thin latex membrane forming a sealed, air-filled cavity together with the generator walls and the liquid film (see Fig. \ref{fig:OscillatorSchematic}a), allowing approximately planar wavefronts to reach the film. Applying an alternating potential to the solenoid causes the magnet, the membrane and, therefore, the liquid film to oscillate.
Using a Laser Doppler vibrometer by Polytec pointed at a piece of TiO\textsubscript{2} foil floating in the center of the liquid film, it is possible to show a linear dependency of the amplitude of the film oscillation on the voltage applied to the oscillator (see Fig. \ref{fig:Vibrometer}a).

All measurements were performed in an aluminum box and with shielded cables.

\section{\label{sec:Materials}Materials}
We are utilizing organic mixed ionic electronic conductors (OMIECs) as electrochemical transducers. In such materials, the density of polarons and hence the electrical conductivity is modulated by the density of anions or cations via a redox chemical reaction. In this work, PEDOT serves as a conjugated electronically conductive material, which is permanently doped by the presence of PF$_6$-anions.

We follow the process of AC electropolymerization\,-\,first proposed by Koizumi \textit{et al.} \cite{koizumi2016electropolymerization, ohira2017synthesis}\,-\,to form the OMIEC PEDOT:PF$_6$ inside the liquid film, resulting in the formation of dendritic polymer FNs. This phenomenon is caused by the limited time of a semi-cycle, allowing growth to only occur in areas with a high anion concentration, as they serve as the dopant for the monomer. Therefore, growth is only seen from the anode, while the monomer is inert at the cathode. Because of the switching of polarity, the fibers grow alternatingly from each electrode and ultimately connect between them. An example of a FN used in this work is shown in Figure \ref{fig:Vibrometer}b. To grow such FNs, the solvent needs to dissolve the monomer and the salt and its electrochemical window has to be larger than the oxidation potential of the monomer \cite{Cucchi2022}. Otherwise, oxidation of the solvent will take place suppressing the polymerization. 
TBAPF$_6$ serves as an oxidizing agent for EDOT and as a dopant for the resulting polymer \cite{PhysRevB.56.3659}. Sulfolane proves useful as a solvent due to its low evaporation rate, yielding a longer lifetime of the films \cite{Martin2021}.
Upon exceeding the oxidation potential of the monomer, a radical monomer cation is formed while the anions of the salt drift to the positively charged electrode, where they form an electrical double layer. The radical then gets neutralized by an anion of the salt and forms longer and heavier structures by reacting with other radical monomers provided by the bulk solution or pre-existing polymer fibers, eventually being too heavy to be soluble and thus depositing at the polymerization site \cite{https://doi.org/10.1002/aelm.202100586}.

\section{\label{sec:R&D}Results and Discussion}
\begin{figure*}[!t]
    \centering
    \includegraphics[width=0.82\textwidth]{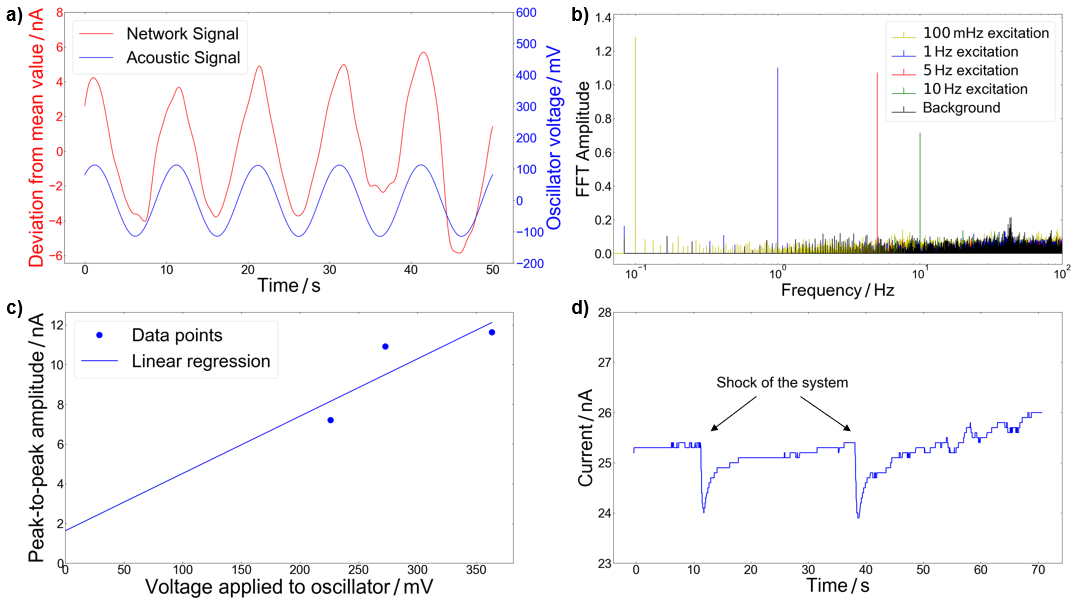}
    \caption{a) External acoustic signal and network response at an excitation frequency of \SI{100}{\milli\hertz}. A digital low-pass filter with a cutoff-frequency twice as large as the respective signal frequency was applied during the data processing. The calculations were performed up to the $250^\text{th}$ order and a Hanning window was used. The peak-to-peak voltage of the acoustic signal is \SI{250}{\milli\volt}. b) FFT of the four measured data sets before the digital low-pass filter was applied, as well as a background measurement without any excitation. c) Mean peak-to-peak amplitude of the current through a FN at an excitation frequency of \SI{500}{\milli\hertz} in dependence on the applied oscillator voltage. For each data point the value was calculated by taking the mean peak-to-peak amplitude of 5 periods. d) Response of a connected FN to a seismic shock caused by dropping an object with a mass of \SI{70}{\gram} on the setup table from a heigt of \SI{15}{\centi\meter}, with a \SI{100}{\milli \volt} DC voltage applied to the electrodes bridged by the FN.}
    \label{fig:4_frequencies}
\end{figure*}

Acoustic excitation of such a system can alter the current that flows through the FN. As shown in Figure \ref{fig:4_frequencies}a, the network signal follows the shape and phase of the sinusoidal excitation potential applied to the oscillator at a frequency of $\SI{100}{\milli\hertz}$. We have two main hypotheses to explain this behavior. Firstly, the reaction might be caused by a change of the doping state in the system. A FN usually grows dendritically with more than one connection (see Fig. \ref{fig:Vibrometer}b). Connected fibers exhibit ohmic behavior, while unconnected fibers exhibit capacitive behavior. During excitation, regions of high and low ion concentration form in the film, causing a slight change in the capacitive properties of the unconnected fibers and thus in the total current flowing through the FN. A periodic excitation hence leads to a network response with the same periodicity. Furthermore, as unconnected fibers are acting as gate electrodes to the connected fibers, a change in capacitance of the unconnected fibers triggers a modulation of the conductivity of the connected fibers \cite{petrauskas2022nonlinear}. In other words, the coupling of the fibers through the electrolyte introduces an internal amplification mechanism to the acoustic excitation.
Secondly, the reaction could have purely mechanical origins. With the liquid film vibrating, fibers might periodically connect and disconnect due to the deformation of the system. This causes a periodic change of the total channel thickness and thus a change of current through the FN dependent on the properties of the acoustic excitation. Even the periodic relative dislocation of the fibers might cause such an effect as the fibers are electrochemically coupled via the electrolyte.

To quantitatively analyze the properties of the system, we performed measurements for four different excitation frequencies with the same applied oscillator voltage of $\SI{250}{\milli\volt}$. 
At $\SI{100}{\milli\hertz}$, the mean peak-to-peak amplitude of the current through the FN for the $5$ periods shown in Figure \ref{fig:4_frequencies}a is $\SI{8.65}{\nano\ampere}$. To calculate the sound pressure generated by the infrasound source, the Young-Laplace equation is used. In sufficiently narrow tubes, the interface forms a meniscus and a factor $2$ has to be integrated into the formula, yielding:

\begin{equation}\label{eq:pressure}
    \Delta p = 2\gamma \cdot \qty(\frac{1}{r_1} + \frac{1}{r_2}),
\end{equation}

where $\gamma$ is the surface tension and $r_1$ and $r_2$ are the principal radii of curvature. The surface tension of sulfolane at room temperature is $\SI{53.3}{\milli\newton\per\meter}$. Since the meniscus is not perfectly spherical, the principal radii of curvature are calculated by approximating the film as a half-ellipsoid with semi-axes lengths of $\SI{2.5}{\milli\meter}$ in x- and y-direction and using the peak-to-peak oscillation amplitude of $\SI{41.37}{\micro\meter}$ at $\SI{100}{\milli\hertz}$ (see Figure \ref{fig:Vibrometer}a) for the semi-axis in z-direction. This gives a value of about $\SI{6.62}{\per\meter}$ for $1/r_1$ and $1/r_2$. Subbing these values into \eqref{eq:pressure} yields a sound pressure of approximately $\SI{1.41}{\pascal}$ and thus a sensitivity of $\SI{613}{\micro\volt\per\pascal}$ at $\SI{100}{\milli\hertz}$ with the used amplification of $10^5\,\si{\volt\per\ampere}$.
During the experiments, a DC voltage of $\SI{300}{\milli\volt}$ is applied to the FN resulting in a baseline current of approximately $\SI{1.12}{\micro\ampere}$ and a power consumption of about $\SI{340}{\nano\watt}$. 
The sensitivity, but especially the low power consumption compare well with other infrasound sensors (see Table \ref{tab:Sensor_Comparison}). However, our system does not require the complex integration of mechanical or optical transducers.
\begin{table}[!t]
\caption{Comparison of Infrasound Sensors.}
\label{tab:Sensor_Comparison}
\centering
\begin{IEEEeqnarraybox}[\IEEEeqnarraystrutmode\IEEEeqnarraystrutsizeadd{1pt}{1pt}][b][\columnwidth]{s+t+t+t}
\IEEEeqnarraydblrulerowcut\\
\IEEEeqnarrayseprow[1pt]{}\\
Frequency range&Sensitivity&Power consumption&Ref.\IEEEeqnarraystrutsizeadd{0pt}{-1pt}\\
\IEEEeqnarrayseprow[1pt]{}\\
\IEEEeqnarrayrulerow\\
\IEEEeqnarrayseprow[1pt]{}\\
0.01-40\,\si{\hertz}&\SI{45.13}{\micro\volt\per\pascal}&\SI{27}{\milli\watt}&\cite{marcillo2012implementation}\\
0.06-40\,\si{\hertz}&\SI{20}{\milli\volt\per\pascal}&\SI{42}{\milli\watt}&\cite{grangeon2019robust}\\
1-20\,\si{\hertz}&\SI{121}{\milli\volt\per\pascal}@\SI{1}{\hertz}&unknown&\cite{wang2016infrasound}\IEEEeqnarraystrutsizeadd{0pt}{-1pt}\\
at least 0.1-10\,\si{\hertz}&\SI{613}{\micro\volt\per\pascal}&\SI{340}{\nano\watt}&our scheme\\
\IEEEeqnarrayseprow[1pt]{}\\
\IEEEeqnarraydblrulerowcut
\end{IEEEeqnarraybox}
\end{table}
A fast fourier transform (FFT) of the network signal was computed (see Figure \ref{fig:4_frequencies}b). The plot does not resemble solely the FN's response, but the gain-bandwidth product due to the integrated low-pass filter, which causes the $\SI{3}{\decibel}$ cutoff frequency at $\SI{10}{\hertz}$. Hence, the bandwidth of the FN itself covers at least the range of $\SI{100}{\milli\hertz}$ to $\SI{10}{\hertz}$.

When changing the voltage applied to the oscillator and hence the oscillation amplitude of the liquid film at a constant excitation frequency, the amplitude of the current flowing through the FN shows an approximately linear behavior (see Fig. \ref{fig:4_frequencies}c). The linearity does not oppose previous works using nonlinear properties of FNs \cite{petrauskas2022nonlinear, cucchi2021reservoir}, since the voltage applied to them is smaller in our case and it thus operates in the linear regime.

The reaction is shown not only for acoustic excitations, but also for seismic shocks of the system. The current flowing through the FN with DC voltage applied drops abruptly upon the impact of the object and then slowly recovers (see Fig. \ref{fig:4_frequencies}d). A possible explanation for the initial drop in current is a mechanical disconnection or weakening of existing fiber connections, before they are able to reconnect due to the liquid film oscillating back to its original position.

\begin{figure}[!t]
    \centering
    \includegraphics[width=0.475\textwidth]{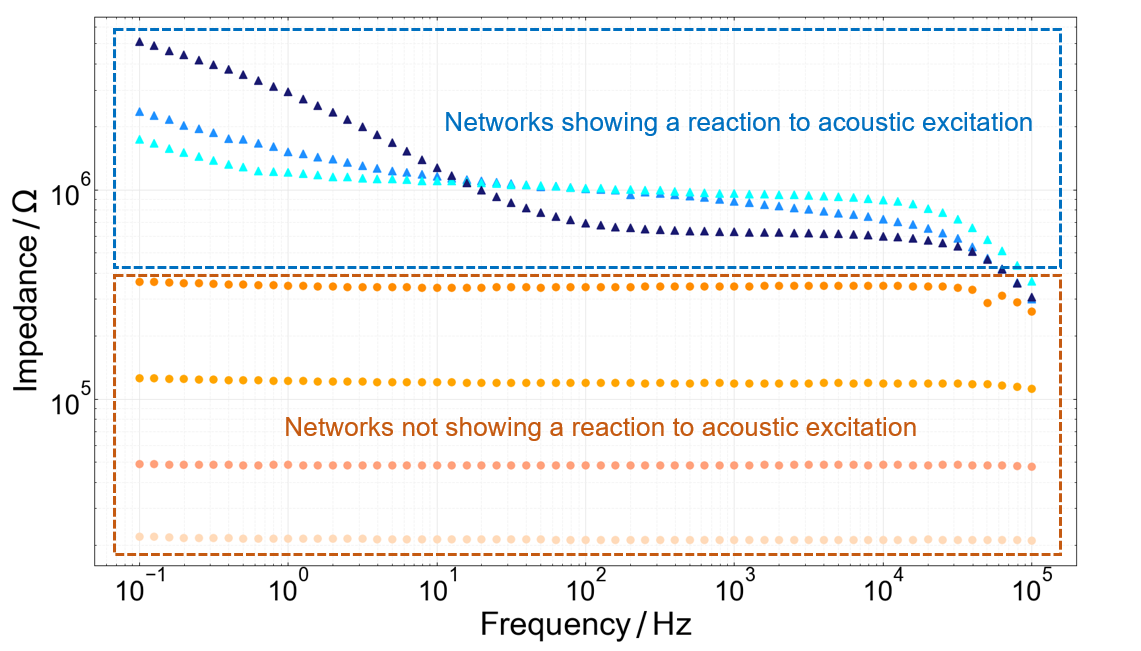}
    \caption{Impedance spectra of PEDOT:PF$_6$ FNs. The spectra with circular markers correspond to purely resistive FNs, the ones with triangular markers correspond to capacitive FNs. The former displayed no current modulation when excited acoustically, the latter did.}
    \label{fig:Impedance}
\end{figure}

The resistive and capacitive properties of PEDOT fibers can vary significantly from one FN to another. Our experiments found that all investigated FNs with capacitive characteristics also showed the current modulation described above when excited acoustically. This reinforces our first hypothesis, that both unconnected and connected fibers are needed to obtain a system with capacitive properties in the low-frequency range. Due to this capacitive response, the polymer fibers can be doped or dedoped in reaction to changes in the local ion concentration when being acoustically or seismically stimulated. FNs with purely ohmic behavior displayed no change in the current passing through them while exposed to infrasound (see Figure \ref{fig:Impedance}).
Unfortunately, we have only limited control over the charge transport properties of the PEDOT fibers. If FNs are grown with low-frequency potentials applied to the gold electrodes, they have more time per cycle to polymerize at the tip of the electrode/fiber and hence grow thicker and exhibit conduction properties similar to an ohmic resistor. Thinner fibers, which grow at higher frequencies, tend to have higher resistances and in some cases also show capacitive behavior. However, high growth frequencies by themselves do not guarantee capacitive properties of the FNs. Currently, we cannot fully control whether a fiber will assume a capacitive behavior. This aspect of the fiber growth process needs further investigations. 

\section{\label{sec:Conclusion}Conclusion}
In this work, we use electropolymerization by applying an AC signal to gold electrodes immersed in free-standing sulfolane films containing EDOT and TBAPF$_6$ to grow dendritic PEDOT:PF$_6$ FNs. The FNs are excited acoustically by an infrasound generator with controllable frequency and strength, causing a modulation of the currents passing through FNs with capacitive properties when a DC-potential is applied. Excitation frequencies are tested between $\SI{100}{\milli\hertz}$ and $\SI{10}{\hertz}$, with a linear response of the FN, making this system a possible candidate for infrasound sensors. The main challenge lies in gaining more control over the growth process to repeatedly achieve a capacitive behavior of the FNs. The infrasound sensor shows comparable properties (sensitivity, bandwidth, and power consumption) to existing device concepts but offers simple integration without complex mechanical or optical transducers.

\end{document}